\documentstyle[12pt,amssymbols]{article}
\input math_macros.tex

\begin{document}
\begin{titlepage}
\today          \hfill 
\begin{center}
\hfill    UCB-PTH-96/30 \\
\hfill    LBNL-39063\\
\vskip .5in

%{\large \bf title}
%\footnote{This work was supported by the Director, Office of Energy 
%Research, Office of High Energy and Nuclear Physics, Division of High 
%Energy Physics of the U.S. Department of Energy under Contract 
%DE-AC03-76SF00098.}
%\vskip .50in

%alternate footnote for faculty:
{\large \bf Topologically nontrivial time-dependent chiral condensates}
\footnote{This work was supported in part by the Director, Office of 
Energy Research, Office of High Energy and Nuclear Physics, Division of 
High Energy Physics of the U.S. Department of Energy under Contract 
DE-AC03-76SF00098 and in part by the National Science Foundation under 
Grant PHY-95-14797.}

\vskip 1.0cm

Mahiko Suzuki

\vskip 0.5cm

{\em Department of Physics\\
           and\\
      Lawrence Berkeley National Laboratory\\
      University of California\\
      Berkeley, California 94720}
\end{center}

\vskip .5in

\begin{abstract}
 Topologically nontrivial time-dependent solutions
of the classical nonlinear $\sigma$ model
are studied as candidates of the disoriented
chiral condensate (DCC) in 3+1 dimensions. Unlike the analytic
solutions so far discussed, these solutions 
cannot be transformed into isospin-uniform ones
by chiral rotations. If they are produced as DCCs, 
they can be detected by a distinct pattern in the angle-rapidity 
distribution of the neutral-to-charged pion ratio.  

\end{abstract}

\vskip 0.3in

PACS: 11.27.+d, \, 11.30.Rd, \, 12.39.Fe, \, 13.85.Hd,\ 
\end{titlepage}
%THIS PAGE (PAGE ii) CONTAINS THE LBL DISCLAIMER
%TEXT SHOULD BEGIN ON NEXT PAGE (PAGE 1)
%\renewcommand{\thepage}{\roman{page}}
%\setcounter{page}{2}
%\mbox{ }

%\vskip 1in

%\begin{center}
%{\bf Disclaimer}
%\end{center}

%\vskip .2in

%\begin{scriptsize}
%\begin{quotation}
%This document was prepared as an account of work sponsored by the United
%States Government. While this document is believed to contain correct 
%information, neither the United States Government nor any agency
%thereof, nor The Regents of the University of California, nor any of their
%employees, makes any warranty, express or implied, or assumes any legal
%liability or responsibility for the accuracy, completeness, or usefulness%
%of any information, apparatus, product, or process disclosed, or represents
%that its use would not infringe privately owned rights.  Reference herein
%to any specific commercial products process, or service by its trade name,
%trademark, manufacturer, or otherwise, does not necessarily constitute or
%imply its endorsement, recommendation, or favoring by the United States
%Government or any agency thereof, or The Regents of the University of
%California.  The views and opinions of authors expressed herein do not
%necessarily state or reflect those of the United States Government or any
%agency thereof, or The Regents of the University of California.
%\end{quotation}
%\end{scriptsize}

%\vskip 2in

%\begin{center}
%\begin{small}
%{\it Lawrence Berkeley Laboratory is an equal opportunity employer.}
%\end{small}
%\end{center}

\newpage
\renewcommand{\thepage}{\arabic{page}}
\setcounter{page}{1}
%THIS IS PAGE 1 (INSERT TEXT OF REPORT HERE)

\section{Introduction}
     Analytic solutions of the classical nonlinear $\sigma$ model
\cite{analytic1, analytic2, analytic3}
have been studied as candidates of the disoriented chiral condensates (DCCs)
\cite {DCC}.
The solutions so far obtained are either configurations with spatially uniform 
isospin distribution or those 
which are chirally equivalent to them. 
When the isospin-uniform DCCs decay, 
the decay pions will obey the event-by-event pion charge
distribution of $dP/df = 1/2\sqrt{f}$ in the neutral pion 
fraction $f$.  In this paper we study as DCC candidates the 
time-dependent solutions of the nonlinear $\sigma$-model 
that are topologically nontrivial in the isospin-orbital space.
Though the topologically nontrivial DCCs obey the same charge 
distribution $dP/df$ as that of statistically random emission,
the angle-rapidity distribution of pions should exhibit a very distinct 
experimental signature.  We suggest a quantitative method of analysis to
search for these DCCs.

\vskip 0.5in

\section{Topologically nontrivial solutions}
    
     Let us express the pion field $\mbox{\boldmath $\pi$}(x)$ 
nonlinearly in terms of the
scalar phase field $\theta(x)$ and the unit isovector field ${\bf n}(x)$ as
\begin{equation}
               \mbox{\boldmath $\pi$}(x)= f_{\pi}{\bf n}(x)\theta(x),
\end{equation}
where $f_{\pi}= 93$ MeV is the pion decay constant. Apart from the source 
term, the Lagrangian is given in terms of $\Sigma(x)=\exp(i{\bf n}(x)\!\cdot
\!\mbox{\boldmath$\tau$}\,\theta(x))$ by
\begin{equation}
    {\cal L}= \frac{f^2_{\pi}}{4}{\rm tr}(\partial_{\mu}\Sigma^\dagger
     \partial^{\mu}\Sigma) + \frac{1}{2}\lambda f_{\pi}^2({\bf n}^2-1),
\end{equation}
in the chiral symmetry limit. After elimination of the Lagrange multiplier 
$\lambda$, the Euler-Lagrange equation reads \cite{analytic2}
\begin{equation}
   \Box\theta-\sin\theta\,\cos\theta\,(\partial_{\mu}{\bf n}\cdot\partial^{\mu}
    {\bf n}) = 0,
\end{equation}
\begin{equation}
    \partial_{\mu}(\sin^2\theta\,\,{\bf n}\times\partial^{\mu}{\bf n}) = 0.
\end{equation}

    The analytic solutions so far known are either those with
${\bf n}(x) = constant$ (the Anselm class) or those which are rotated to them
by chiral transformations.  
In this paper we explore the class of solutions whose ${\bf n}(x)$ 
fields point radially in the spacetime of $3+1$ dimensions:
\begin{equation}
                  {\bf n}(x) = \frac{{\bf r}}{r}\,.
\end{equation}
We have obtained a hint for this postulate from the 
Skyrme assumption that led to
the static soliton \cite{skyrme}. Unlike Skyrme, we do not 
need a stabilizing term in the Lagrangian since the 
static stability of solutions
is irrelevant to us.  The spherical symmetry of ${\bf n}(x)$ suggests 
we should choose $\theta(x)$ also to be spherically symmetric as 
\begin{equation}
                     \theta(x) =\theta(t, r).
\end{equation}
It is important to observe that with our postulate the Euler-Lagrange equation 
(4) for ${\bf n}(x)$ is automatically satisfied for any $\theta(t,r)$.  
The equation for $\theta(x)$ now reads
\begin{equation}
       (\partial^2_t-\nabla^2)\theta(t,r)+\frac{2}{r^2}\sin\theta(t,r)
        \,\cos\theta(t,r)=0.
\end{equation}
This wave equation allows many interesting solutions. All of them are
topologically nontrivial since following the Skyrme model we can introduce
the topological charge,
\begin{equation}
             q = \int Q_0 d^3x,
\end{equation}
where    
\begin{equation}
 Q_{\mu}(x) = \frac{\epsilon_{\mu\nu\kappa\lambda}}{24\pi^2}
              {\rm tr}(X^{\nu}X^{\kappa}X^{\lambda}),
\end{equation}
with $X^{\nu} = \Sigma^\dagger\partial^{\nu}\Sigma$. The current $Q_{\mu}$ 
is locally conserved, $\partial^{\mu}Q_{\mu}=0$, 
and $q$ is invariant under chiral $SU(2)\times SU(2)$ rotations. 
The charge $q$ is nonvanishing for our solutions while it 
is zero for the isospin uniform solutions.
It should be noted that $q$ is time dependent when we compute it
for the  pion fields alone
since the current $Q_{\mu}$ flows from the shell of hadron 
debris into the DCC.  Actually it is not even finite since our 
pion fields are singular as we approach the light cone.  This should not 
bother us since the nonlinear $\sigma$ model after all does not apply
to the close neighborhood of the light cone where kinetic energy
is too large. Our purpose of mentioning the topological
charge $q$ here is that our solutions are chirally inequivalent 
to the isospin-uniform solutions. 
When we adopt our solutions as DCC candidates, 
we do not accept the Skyrme model of baryons \cite{nappi} in which 
the charge $q$ is identified with the baryon number.  
If we did, our DCCs would be loaded with nucleons or antinucleons.

Let us solve for $\theta(t,r)$ by restricting
the form of $\theta(t,r)$. Since $\theta$ is dimensionless and its equation 
of motion is scale invariant, a simple case of interest is that
$\theta(t,r)$ is a function only of the ratio of $r$ and $t$:
\begin{equation}
          \theta(t,r) = \theta(\xi), \;\;\;\;\; (\xi \equiv\frac{r}{t} ).  
\end{equation}
The wave equation then turns into
\begin{equation}                  
    \frac{d^2}{d\xi^2}\theta + \frac{2}{\xi}\theta
        - \frac{2}{\xi^2(1-\xi^2)}\sin\theta\,\cos\theta =0. \label{theta'}
\end{equation}
It can be easily solved numerically. The behavior at $\xi=0$ is determined 
by the singularity at $\xi=0$ in the wave equation (\ref{theta'}).  
Barring a singularity at $r=0$ for $\theta(x)$ 
since a source does not exists at $r=0$ after $t=0$, we
determine that $\theta(\xi)\propto\xi$ as $\xi\rightarrow 0$. By giving one 
more boundary condition, $d\theta/d\xi$ at $\xi=0$, we can compute a profile of
the scalar phase field $\theta(\xi)$.  In Fig.1 we have plotted $\theta(\xi)$
for a few different values of $\theta'(0)$. On the light cone ($\xi=1$),
$\theta(t,r)$ is singular because of $1/(1-\xi^2)$ in the third term of
Eq.(\ref{theta'}), but only in the immediate neighborhood of $\xi=1$.
The function $\theta(\xi)$ is smooth and monotonic
practically everywhere inside the light cone.
           
When $\theta(t,r)$ is not a function only of $\xi=r/t$, analyticity at $r=0$
requires the behavior $\theta(t,r)\rightarrow r$ at $r\rightarrow 0$,
not necessarily $\theta(t,r)\rightarrow r/t$. We have drawn the asymptotic
configuration $\theta(\infty, r)$ in Fig.2.  The static equation for
$\theta(\infty, r)$ is invariant under rescaling of $r$. As we know from the
Skyrme model, this $\theta(\infty, r)$ is not a local minimum of 
energy with respect to rescaling of $r$ since there is no Skyrme term in 
our Lagrangian.  In the dynamical case under considerations, solutions of all
different scales are acceptable for $\theta(\infty, r)$.  
The amount of energy fed
in by the hadron debris determines the scale of a DCC that is produced.

From the viewpoint of total energy, the nontrivial {\bf n}(x) costs nothing
since $\Sigma(x)$ remains at the bottom of the double-well potential and
the static ${\bf n}({\bf r})$ does not contribute to kinetic energy.  As far
as energy is concerned, the topologically nontrivial solutions are no different
from the isospin-uniform solutions. 

\vskip 0.5in

\section{Isospin property}
     It might appear that the topologically nontrivial DCCs have zero isospin
because of the {\it spherical symmetry}.  It is not correct.
At the quantum level they are not eigenstates of $I=0$ but superpositions of 
eigenstates of many different isospin values.  
To see it, we should represent the quantum state of the classical 
configuration by the coherent state \cite {glauber, suzuki}:
\begin{equation}
  |{\bf n}({\bf r})\theta(t,r) \rangle =
  \exp\biggl(-if_{\pi}
  \int{\rm tr}\Bigl(\theta(t,r){\bf n}({\bf r})\cdot\partial_t 
  \mbox{\boldmath$\pi$}(x)-\partial_t\theta(t,r)\,{\bf n}({\bf r})\cdot
  \mbox{\boldmath$\pi$}(x)\Bigr)d^3x\biggr)|0\rangle,  \label{state}
\end{equation}
where $\mbox{\boldmath$\pi$}(x)$ is the isovector quantum pion field.  
The exponent is invariant under the simultaneous isospin-orbital 
rotations generated by ${\bf K}={\bf I} + {\bf L}$, where {\bf L} is orbital
angular momentum. Therefore this state is an eigenstate of $K=0$,
not of $I=0$ nor $I_3 =0$ \cite{jackiw}.  The exponent of Eq.(\ref{state}) 
can be expressed in terms of the creation operators 
$a_{klm}^{(\alpha)\dagger}$ 
of pions with charge $\alpha = (+, -, 0)$, energy $k(\equiv |{\bf k}|)$,
and orbital angular momentum ($l, m$) as
\begin{equation}
      \propto \int\Bigl(a^{(-)\dagger}_{k11}
    +a^{(0)\dagger}_{k10}+a^{(+)\dagger}_{k1\,-1}\Bigr)
    \sqrt{2k}\,\tilde{\theta}_{l}(k)kdk \;-\; h.c.,
                 \label{exponent}
\end{equation}
where $\tilde{\theta}_{l}(k)$ is the Bessel transform 
$\int\sqrt{kr} J_{l+1/2}(kr)\theta(0,r)r\,dr$.
Projection onto the $N_{\pi}$ pion state of Eq.(\ref{state}) is
$N_{\pi}$-th power of Eq.(\ref{exponent}) operated on $|0\rangle$.
By isospin decomposition we see that the state is not purely an isosinglet 
but has a wide distribution in $I$. It is remarked that the initial state 
of $p\bar{p}$ collision has $K=0$ or $1$ 
in the case of head-on collision with zero impact
parameter since $L=0$ and $I=0$ or $1$.
 
%Since we have been unable 
%to obtain analytic form, we make a qualitative argument.
%By commutativiy of $a^(i)\dagger(k)$ for different $i=(1,2,3)$, we 
%can write the exponential operatorof the coherent state (\ref{state}) 
%in product of three factors.  Then the distribution (\ref{isospin}) holds for
%each factor.  The isospin distribution for the product of three factors is
%basically the same as each.  Therefore we expect that despite the spherical
%spherical symmetry in isospin and orbital the topologically nontrivial DCCs
%will also be suffered from the same isospin suppression as the isospin uniform
%DCCs.

%   From the viewpoint of the total DCC energy, the nontrivial 
%${\bf n}({\bf r})$ costs nothing since $\Sigma(x)$ remains at the 
%bottom of the double-well potential 
%and the static ${\bf n}({\bf r})$ does not contribute to kinetic
%energy.  As far as energy is concerned, the topologically nontrivial solutions 
%are no different from the isospin-uniform solutions.

\vskip 0.5in

\section{Momentum distribution of neutral-to-charged pion ratio}
     We study the spectrum of the pions decaying from the topologically
nontrivial DCCs.  We focus on the correlation between the isospin 
and momentum distributions since it shows a distinct characteristic. 
Let us describe the expanding hadron debris
of the baked Alaska scenario \cite{Alaska} by the isovector source 
$\mbox{\boldmath$\rho$}(x)$ ($=\Box\mbox{\boldmath$\pi$}(x)$).
The standard method \cite{Henley} gives
the momentum spectrum of the pions with the Cartesian 
isospin component $i$ and momentum ${\bf k}$ as
\begin{equation}
 2k_0\frac{dN_{i}}{d^3{\bf k}}=\frac{1}{(2{\pi})^3}|\tilde{\mbox{\boldmath
         $\rho$}}({\bf k})\cdot{\bf e}_i|^2,   \label{distribution}
\end{equation}
where $\tilde{\mbox{\boldmath$\rho$}}(\bf k)$ is the four-dimensional
Fourier transform with $k_0 = |{\bf k}|$ for massless pions,
\begin{equation}
   \tilde{\mbox{\boldmath$\rho$}}({\bf k}) 
      = \int\mbox{\boldmath$\rho$}(x)\, e^{ikx} d^4 x,
\end{equation}    
and ${\bf e}_i$ is a unit Cartesian isospin vector along the {\it i}-th
direction. The spherically symmetric pion fields, $\theta(t,r)$ and 
${\bf n}(x)= {\bf r}/r$, can be produced only by  
$\mbox{\boldmath$\rho$}(x)$ of the same symmetry.  Therefore 
$\mbox{\boldmath$\rho$}(x)$ may be expressed as 
\begin{equation}
      \mbox{\boldmath$\rho$}(x) = 
         \frac{{\bf r}}{r}\rho(r)\delta(t^2-r^2).
\end{equation} 
Then the ${\bf k}$ dependence of its
Fourier transform is determined in the form
\begin{equation}
    \tilde{\mbox{\boldmath$\rho$}}({\bf k})
      = \frac{{\bf k}}{|{\bf k}|}\tilde{\rho}(|{\bf k}|). \label{rho}
\end{equation} 
Substituting Eq.(\ref{rho}) in Eq.(\ref{distribution}), 
we find that the pion isospin is correlated
with the momentum direction as
\begin{equation}
          2k_0 \frac{dN_i}{d^3{\bf k}} \propto ({\bf k}\cdot{\bf e}_i)^2.
               \label{correlation}
\end{equation}
Namely, neutral pions are emitted preferentially along the 
$z$-axis while charged pions are into the $xy$-plane.  
Note however that the $z$-axis does not necessarily coincide with the
collision axis. For some DCC, the z-axis 
happens to be the collision axis. Then many other DCCs can exist
that are related to this one by isospin rotations.  They form 
one isospin family of DCC solutions.  When the $z$-axis coincides with
the collision axis, we expect to see a pair of parallel $\pi^0$-rich bands
around $y_1$ and $y_2$ in the $\phi$-$y$ plot, where
$\phi$ is the azimuthal angle of the pion momentum 
${\bf k}$ around the collision axis and $y$ is the rapidity variable.
The region between the two bands is filled
dominantly with charged pions.  When the isospin axis 
is not parallel to the collision axis, 
a pair of $\pi^0$-rich domains is found at ($\phi$, $y_1$) 
and ($\phi + \pi$, $y_2$) in the $\phi$-$y$ plot.  
The two $\pi^0$-rich domains are separated by a $\pi^{\pm}$-rich region
(Fig.3). 

The event-by-event distribution $dP/df$ of the neutral pion fraction 
$f=N_{\pi^0}/(N_{\pi^+}+N_{\pi^-}+N_{\pi^0})$ 
can be obtained from Eq.(\ref{correlation}).
By integrating over the angles of ${\bf k}$,
we find that the particle multiplicity is
equal for all three pion charge states. 
Therefore the distribution $dP/df$
cannot distinguish the topologically nontrivial DCC events from
random emission events. 

We can make search of the topologically nontrivial DCCs more quantitative.   
We should first select the DCC candidates, for instance, by
abundance of soft $p_t$ pions and find their approximate overall
rest frames. To enhance the signature, we should select only those events
with $f\approx 1/3$. We then determine event by event the isospin {\it polar} 
direction $\hat{{\bf z}}$, namely the $\pi^0$-dominant
direction in the momentum space, by maximizing a suitably defined 
quantity as follows:  By choosing a tentative $\hat{z}$
direction, compute for ${\pi^0}$ and ${\pi^{\pm}}$
\begin{eqnarray}
   C_0 & = & \sum_{i=\pi^0}(\hat{{\bf k}}_i\cdot\hat{{\bf z}})^2,  \nonumber \\
   C_{\pm}&=&\sum_{i=\pi^{\pm}}(\hat{{\bf k}_i}\times\hat{{\bf z}})^2, 
\end{eqnarray}
where $\hat{{\bf k}}_{0}$ and $\hat{{\bf k}}_{\pm}$ 
are the unit vectors along neutral and charged pion momenta, respectively, 
and the summations are taken over all DCC pions from each event ($N_{\pi^0}
\approx N_{\pi^+}\approx N_{\pi^-}$). Then vary the direction 
of $\hat{z}$ so as to maximize the product 
\begin{equation}
   C  = C_0\, C_{\pm}.
\end{equation}
The quantity $C$ takes the maximum value when $\hat{z}$ 
coincides with the isospin polar axis.
Let this maximum be $C_{max}$.  Then compute $C$ for the same 
$\hat{z}$ direction by interchanging $\pi^{\pm}$ and $\pi^0$. 
Let us call it $C_{min}$.  Then the ratio
\begin{equation}
       S =\frac{C_{max} - C_{min}}{C_{max}+ C_{min}} 
\end{equation}
is equal to 5/7 for the topologically nontrivial DCCs 
with sufficiently large $N_{\pi}$. As the direction of $\hat{z}$ is varied,
$S$ sweeps between $5/7$ and $-5/7$.  In contrast, $S$ is independent of
the direction of $\hat{z}$ and equal to zero for the isospin-uniform
DCCs as well as for random emission.  
Feasibility of this test depends on how large $N_{\pi}$ is.  
Since the statistical errors are of O$(1/\sqrt{N_{\pi}})$, 
$N_{\pi}\,\gtap\, 25$ will do.

\vskip 0.5in

\section{Chiral rotations and disorientation}
    The topologically nontrivial solutions of the form
$\theta(\xi)$ behave like 
$\sim\xi$ at $\xi (= r/t)\rightarrow 0$.  At sufficiently large $t$, 
therefore, the phase
pion field $\theta(\xi)$ approaches zero, that is, the background vacuum
relaxes to the true vacuum at any fixed location off the light cone. 
If the DCC is defined to be the disoriented
condensate that would approach a {\it wrong} vacuum at $t\rightarrow\infty$, 
one might not want to call such a condensate as the DCC \cite{analytic3}.  
When we make chiral rotations on $\theta(\xi)$, we obtain
infinitely many more solutions that carry the same topological charge
but different vector-isospin distributions. 
By actually performing the axial-isospin rotations, 
we can obtain solutions whose ${\bf n}(x)$ is not spherical
at finite $t$.  As $t\rightarrow\infty$, these rotated 
solutions approach static limits with uniform isospin orientations 
at all finite locations:
\begin{equation}
       \lim_{t\rightarrow\infty}{\bf n}(x) \theta(x) = {\bf n}_0\theta_0, 
\end{equation}
where ${\bf n}_0$ is the axis of an axial-vector isospin rotation (${\bf n}_L
= {\bf n}_R \equiv {\bf n}_0$) and $\theta_0$ is its rotation angle 
($\theta_L = -\theta_R \equiv \theta_0$) \cite{analytic2,analytic3}. 
Namely, the asymptotic disorientation stays nonvanishing and turns uniform. 

When $\theta(t,r)$ is not a function only of $\xi= r/t$,
we have found $\theta(t,r)\rightarrow r$ at $r\rightarrow 0$, instead of
$\rightarrow r/t$. It remains {\it disoriented} and {\it nontrivial} as
${\bf n}(x) = {\bf r}/r$ even at $t=\infty$.  Chiral rotations transform
${\bf n}(x)={\bf r}/r$ into nonspherical ${\bf n}(x)$.

\vskip 0.5in

\section{Explicit chiral symmetry braking}

   In the presence of an explicit chiral symmetry breaking, 
a DCC cannot really reach its asymptotic limit predicted by
the chiral symmetric wave equations. It starts decaying
before its kinetic energy $f_{\pi}^2\dot{\theta}^2/2$ becomes 
comparable with the potential difference of the symmetry breaking
$m_{\pi}^2 f_{\pi}^2 (1-\cos\theta)$. 
However, the postulate of ${\bf n}(x)= {\bf r}/r$ 
is still compatible with the explicit chiral symmetry breaking.
Furthermore the topological charge $q$ is
defined in the same form and the local current conservation 
$\partial^{\mu}Q_{\mu}=0$ remains valid. 
The only change is appearance of a scale breaking term 
proportional to $m_{\pi}^2$ in the wave equation of $\theta(t,r)$: 
\begin{equation}
 \Box\theta + \frac{2}{r^2}\sin\theta\,\cos\theta = m_{\pi}^2\sin\theta. 
\end{equation} 
With the explicitly scale dependent term present, we can no longer 
postulate $\theta=\theta(\xi)$. Instead $\theta(x)$ is a function of two
dimensionless variables $r/m_{\pi}$ and $t/m_{\pi}$.
But the wave equation still allows a spherically symmetric
$\theta(t,r)$, and the behavior
$\theta(t,r)\sim r$ at $r\rightarrow 0$ is not affected by the symmetry 
breaking. As long as the form of ${\bf n}({\bf r})$ is the same, 
our prediction in the $\phi$-$y$ distribution
of the decaying pions is valid with no modification.
 
\vskip 0.5in

\section{Conclusion}
   We have argued that topologically nontrivial DCCs are not only possible but
also have an equally good or bad change of being produced as 
the isospin-uniform ones. If they are actually produced, they
will show a clear experimental signature in the $\phi$-$y$ plot of the decay
pions. In order to produce them the hadron debris
must also expand with a spherically symmetric isospin distribution
${\bf J}_0\propto \hat{{\bf r}}$ according to geometrical symmetry. 
No dynamical mechanism is known that 
suppresses such a configuration of hadron debris.  
In a DCC search the topologically nontrivial 
DCCs should also be searched for by the analysis proposed here.

\vskip 0.5in

\section{Acknowledgment} 
   This work was supported in part by the National Science Foundation under 
Grant PHY-95-14797 and in part by the Director, Office of Energy Research,
Office of High Energy and Nuclear Physics, Division of High Energy Physics of
the U.S. Department of Energy under Contract DE-AC03-76SF00098.

\newpage
{\bf  Figure captions}

\vskip 0.5in

   Fig.1: The scalar phase function $\theta(\xi)$ for a few different
boundary values of $\theta'(0)$. $\theta(\xi)\propto\xi$ is required at
$\xi\rightarrow 0$ by the wave equation.   

\vskip 0.3in

   Fig.2: The asymptotic configuration of $\theta(t,r)$ at $t=\infty$ in the
case that $\theta(t,r)$ is not a function only of $\xi = r/t$. 
The variable $r$ is expressed in the unit of $m_{\pi}$. 

\vskip 0.3in 

   Fig.3: Schematic pictures of the pion charge distribution in the 
$\phi$-$y$ plot for a topologically nontrivial DCC   
whose $I_3$ direction is along the collision axis (3a)
and off the collision axis (3b).

%\vskip 0.3in

%   Fig.3: The asymptotic configuration of $\theta(t,r)$ at $t=\infty$ 
%in the case that $\theta(t,r)$ is not a function only of $\xi=r/t$.  


\vskip 0.5in

\begin{thebibliography}{99}
\bibitem{analytic1} J.-P. Blaizot and A. Krzywicki, Phys. Rev. D{\bf 46}, 246
     (1992).
\bibitem{analytic2} Z. Huang and M. Suzuki, Phys. Rev. D{\bf 53}, 891 (1996).
\bibitem{analytic3} M. Suzuki, LBNL38931/UCB-PTH-96/23/hep-ph/9606234, to 
     be published in Phys. Rev. D.
\bibitem{DCC}  A. A. Anselm, Phys. Lett. B{\bf 217}, 169 (1989),\\
     A. A. Anselm and M. G. Ryskin, Phys. Lett. B{\bf 266}, 482 (1991),\\
     J. D. Bjorken, Int. J. Mod. Phys. A{\bf 7} 4189 (1992); Acta Phys. Pol.
     B{\bf 23}, 561 (1992),\\
     J. D. Bjorken, K. L. Kowalski, and C. C. Taylor, SLAC Report SLAC-PUB-6109
     (1992), unpublished,\\
     K. L. Kawalski nad C. C. Taylor, Case Western Reserve Univerity Report
     CWRURTH-92-6 (1992), unpublished,\\
     K. Rajagopal and F. Wilczek, Nucl. Phys. B{\bf 399}, 395 (1992); 
     B{\bf 404}, 577 (1993),\\
     Z. Huang and X.-N. Wang, Phys. Rev. D{\bf 49}, R4339 (1994),\\
     M. Asakawa, Z. Huang, and X.-N. Wang, Phys. Rev. Lett. {\bf 74},
     3126 (1995),     \\
     F. Yu. Klebnikov, Mod. Phys. Lett. A{\bf 8}, 1901 (1993),\\
     A. Krzywicki, Phys. Rev. D{\bf 48}, 5190 (1993),\\
     A. A. Anselm and M. Bander, Pis'ma Zh. Eskp. Teor. Fiz.{\bf 59}, 479
     (1994),\\
     I. I. Kagan, Phys. Rev. D{\bf 48}, R3971 (1993); Pis'ma Zh. Eksp. Teor. 
     Fiz. {\bf 59}, 289 (1994)[JETP Lett. {\bf 59}, 307 (1994)],\\
     S. Gavin, A.Gocksch, and R. D. Pisarski, Phys. Rev. Lett.{\bf 72}, 2143
     (1994),\\
     S. Gavin and B. Muller, Phys. Lett. {\bf 329}B, 486 (1994),\\
     Z. Huang and X.-N. Wang, Phys. Rev. D{\bf 49}, R4339 (1994),\\
     M. Asakawa, Z. Huang, and X.-N. Wang, Phys. Rev. Lett. {\bf 74}, 3126
     (1995),\\
     R. Amado and I. I. Kagan, Phys. Rev. D{\bf 51}, 190 (1995),\\ 
     D. Boyanovsky, H. J. de Vega, and R, Holman, Phys. Rev. D{\bf 51}, 734
     (1995),\\
     F. Cooper, Y. Kluger, E. Mottola, and J. P. Paz, Phys. Rev. D{\bf 51}, 2377
     (1995),\\
     A. Bialas, W. Cryz, and M. Gmyrek, Phys. Rev. D{\bf 51}, 3239 (1995),\\
     A. Barducci, L. Caiani, R. Cassalbuoni, M. Mondugnio, G. Pettini, and R.
     Gatto, Phys. Lett. {\bf 369}B, 23 (1996),\\
     M. A. Lamper, J. F. Dawson, and F. Cooper, hep-ph/9603668,\\
     J. Randrup, LBL-38379/hep-ph/9605223,\\
     F. Cooper, Y. Kluger, and E. Mottola, LBL-38585.
\bibitem{skyrme}T. H. R. Skyrme, Proc. Roy. Soc. London A{\bf 260}, 127 (1961).
\bibitem{nappi}  A. P. Balachandran, V. P. Nair, S. G. Rajeev, and A. Stern,
     Phys. Rev. D{\bf 27}, 1153 (1983),\\
     G. Adkins, C. Nappi, and E. Witten, Nucl. Phys. B{\bf 228}, 552 (1983).
\bibitem{glauber} R. J. Glauber, Phys. Rev. {\bf 131}, 2766  (1963).
\bibitem{suzuki}  M. Suzuki, Pys. Rev. D{\bf 52}, 982 (1995). The linear 
     $\sigma$ model is used there.
\bibitem{jackiw} R. Jackiw and C. Rebbi, Phys. Rev. Lett.{\bf 36}, 1116 (1976);
    Phys. Rev. D{\bf 13}, 3398 (1976).
\bibitem{Alaska} J. D. Bjorken, K. L. Kowalski, and C. C. Taylor,
     SLAC Report SLAC-PUB-6190, unpublished.
\bibitem{Henley}  E. M. Henley and W. Thirring, {\it Elementary Quantum FIeld
     Theory} (McGraw Hill, New York, 1962), Chapter 8-10.
\end{thebibliography}
\end{document}